\documentclass[twocolumn]{aa} 
\usepackage{graphicx,longtable}
\usepackage[varg]{txfonts}
\begin{document}

   \title{The Optical SN\,2012bz Associated with the Long GRB\,120422A\thanks{Based on observations made with the Italian 3.6-m Telescopio Nazionale Galileo (TNG), operated on the island of La Palma by the Fundaci\'{o}n Galileo Galilei of the INAF (Instituto Nazionale di Astrofisica) at the Spanish Observatorio del Roque de los Muchachos of the Instituto de Astrof\'{i}sica de Canarias under program A25TAC$\_$11, and with the ESO 8.2-m Very Large Telescope (VLT) at Paranal Observatory under program 089.D-0033(A).}}

  \author{A. Melandri\inst{1}, E. Pian\inst{2,3,4}, P. Ferrero\inst{5,6}, V. D'Elia\inst{7,8}, E. S. Walker\inst{2}, G. Ghirlanda\inst{1}, S. Covino\inst{1}, L. Amati\inst{9}, P. D'Avanzo\inst{1},  P. A. Mazzali\inst{10,11}, M. Della Valle\inst{12,13}, C. Guidorzi\inst{14}, L. A. Antonelli\inst{8}, M. G. Bernardini\inst{1}, D. Bersier\inst{15}, F. Bufano\inst{16}, S. Campana\inst{1}, A. J. Castro-Tirado\inst{17}, G. Chincarini\inst{1},  J. Deng\inst{18}, A. V. Filippenko\inst{19}, D. Fugazza\inst{1},  G. Ghisellini\inst{1}, C. Kouveliotou\inst{20}, K. Maeda\inst{21}, G. Marconi\inst{22}, N. Masetti\inst{9},  K. Nomoto\inst{21}, E. Palazzi\inst{9}, F. Patat\inst{23}, S. Piranomonte\inst{8}, R. Salvaterra\inst{24}, I. Saviane\inst{21}, R. L. C. Starling\inst{25}, G. Tagliaferri\inst{1}, M. Tanaka\inst{26}, S. D. Vergani\inst{1,27}
          }

   \institute{$^{1}$ INAF-Brera Astronomical Observatory, via E. Bianchi 46, I-23807 Merate (LC), Italy\\
              \email{andrea.melandri@brera.inaf.it}\\
    {$^{2}$ Scuola Normale Superiore, Piazza dei Cavalieri 7, I-56126 Pisa, Italy}\\
    {$^{3}$ INAF-Trieste Astronomical Observatory, via G.B. Tiepolo 11, I-34143 Trieste, Italy}\\
    {$^{4}$ INFN, Sezione di Pisa, Largo Pontecorvo 3, I-56127 Pisa, Italy}\\
    {$^{5}$ Instituto de Astrof\'{i}sica de Canarias, E-38200 La Laguna, Tenerife, Spain}\\
    {$^{6}$ Departamento de Astrofisica, Universidad de La Laguna, E-38206 La Laguna, Tenerife, Spain}\\
    {$^{7}$ ASI, Science Data Centre, via Galileo Galilei, I-00044 Frascati, Italy}\\
    {$^{8}$ INAF-Roma Astronomical Observatory, via Frascati 33, I-00040 Monteporzio Catone, Italy}\\
    {$^{9}$ INAF, IASF Bologna, via P. Gobetti 101, I-40129 Bologna, Italy}\\
    {$^{10}$ Max Planck Institute for Astrophysics, Karl-Schwarzschild-Strasse 1, D-85748 Garching, Germany}\\
    {$^{11}$ INAF-Padova Astronomical Observatory, vicolo dell'Osservatorio 5, I-35122 Padova, Italy}\\
    {$^{12}$ INAF-Capodimonte Astronomical Observatory, Salita Moiariello 16, I-80131 Napoli, Italy }\\
    {$^{13}$ ICRANET, Piazza della Republica 10, Pescara, I-65122, Italy}\\    
    {$^{14}$ Physics Department, University of Ferrara, via Saragat 1, I-44122, Ferrara, Italy}\\
    {$^{15}$ ARI, Liverpool John Moores University, Twelve Quays House, Egerton Wharf, Birkenhead, CH41 1LD, UK}\\
    {$^{16}$ Departamento de Ciencias Fisicas, Universidad Andr\'{e}s Bello, Av. Rep\'{u}blica 252, Santiago, Chile}\\
    {$^{17}$ Instituto de Astrof\'{i}sica de Andaluc\'{i}a, CSIC, Glorieta de la Astronom\'ia s/n, 18.008 Granada, Spain}\\
    {$^{18}$ National Astronomical Observatories, CAS - Beijing 100012, China}\\
    {$^{19}$ Department of Astronomy, University of California, Berkeley, CA 94720-3411, USA}\\    
    {$^{20}$ NASA Marshall Space Flight Center, Huntsville, Alabama 35805, USA}\\
    {$^{21}$ Kavli Institute for the Physics and Mathematics of the Universe, University of Tokyo, Kashiwa, Chiba 277-8583, Japan}\\
    {$^{22}$ European Southern Observatory, Alonso de Cordova 3107, Santiago, Chile}\\
    {$^{23}$ European Southern Observatory, Karl-Schwarzschild-Str. 2, D-85748, Garching, Germany}\\
    {$^{24}$ INAF, IASF Milano, via E. Bassini 15, I-20133 Milano, Italy}\\
    {$^{25}$ Department of Physics and Astronomy, University of Leicester, University Road, Leicester LE1 7RH, UK}\\
    {$^{26}$ National Astronomical Observatory of Japan, Mitaka, Tokyo 181-8588, Japan}\\
    {$^{27}$ GEPI, Observatoire de Paris, CNRS, Univ. Paris Diderot, 5 pl. Jules Janssen, 92190 Meudon, France}\\   
    }

   \date{}

  \abstract
   {}
   {The association of Type Ic supernovae (SNe) with long-duration gamma-ray bursts (GRBs) is well established. We endeavor, through accurate ground-based observational campaigns, to characterize these SNe at increasingly high redshifts.}
   {We obtained a series of optical photometric and spectroscopic observations of the Type Ic SN\,2012bz associated with the Swift long-duration GRB\,120422A (redshift $z$ = 0.283) using the 3.6-m TNG and the 8.2-m VLT telescopes during the time interval between 4 and 36 days after the burst.}
   {The peak times of the light curves of SN\,2012bz in various optical filters differ, with the $B$-band and $i'$-band light curves reaching maximum at $9 \pm 4$ and $23 \pm 3$ rest-frame days, respectively. The bolometric light curve has been derived from individual bands photometric measurements, but no correction for the unknown contribution in the near-infrared (probably around 10--15\%) has been applied. Therefore, the present light curve should be considered as a lower limit to the actual UV-optical-IR bolometric light curve. This pseudo-bolometric curve reaches its maximum ($M_{\rm bol} = -18.56 \pm 0.06$) at $13 \pm 1$ rest-frame days; it is similar in shape and luminosity to the bolometric light curves of the SNe associated with $z < 0.2$ GRBs and more luminous than those of SNe associated with X-ray flashes (XRFs). A comparison with the model generated for the bolometric light  curve of SN\,2003dh suggests that SN\,2012bz produced only about 15$\%$ less $^{56}$Ni than SN\,2003dh, about $0.35\,{\rm M}_{\odot}$. Similarly the VLT spectra of SN\,2012bz, after correction for Galactic extinction and for the contribution of the host galaxy, suggest comparable explosion parameters with those observed in SN\,2003dh ($E_{\rm K} \sim 3.5 \times 10^{52}$ erg, $M_{\rm ej} \sim 7 M_{\odot}$) and a similar progenitor mass ($\sim 25$--40\,M$_{\odot}$). GRB\,120422A is consistent with the $E_{\rm peak}-E_{\rm iso}$  and the $E_{\rm X,iso}-E_{\rm \gamma,iso}-E_{\rm peak}$ relations. GRB\,120422A\,/\,SN\,2012bz shows the GRB-SN connection at the highest redshift so far accurately monitored both photometrically and spectroscopically.}
   {}

   \keywords{Gamma-ray burst: general; supernovae: individual: SN\,2012bz}

\authorrunning{A. Melandri et al.}
\titlerunning{SN\,2012bz Associated with GRB\,120422A}

   \maketitle
%

\section{Introduction}

Many known X-ray flashes (XRFs) and under-energetic long-duration gamma-ray bursts (GRBs; $E_{\rm iso} < 10^{51}$ erg) have redshifts $z \lesssim 0.25$, and nearly all of them have a well-studied associated supernova (SN) whose detection may have been favored by both the low redshift and the weak luminosity of the optical afterglow (Malesani et al. 2004; Gal-Yam et al. 2004; Mirabal et al. 2006; Modjaz et al. 2006; Pian et al. 2006; Ferrero et al. 2006; Chornock et al. 2010; Cano et al. 2011a; Bufano et al. 2012; Starling et al. 2011; Sollerman et al. 2006; Galama et al. 1998).\footnote{There are two exceptions where the presence of a SN could be excluded to a very deep limit in the maximum luminosity (Della Valle et al. 2006; Fynbo et al. 2006; Gal-Yam et al. 2006).} Among nearby GRBs with a well-monitored SN, only GRB\,030329 (associated with SN\,2003dh; Hjorth et al. 2003; Stanek et al. 2003; Matheson et al. 2003) has an isotropic energy larger than $10^{51}$ erg (Vanderspek et al. 2004; Lipkin et al. 2004), typical of higher redshift, so-called ``classical'' or ``cosmological'' GRBs. 

Low-redshift GRB- and XRF-associated SNe were accurately monitored both photometrically and spectroscopically, and identified as highly stripped-envelope core-collapse SNe (i.e., with no hydrogen and helium), generally called Type Ic SNe (see Filippenko 1997 for a review of SN classification).  They stand apart for their high kinetic energies, up to $5 \times 10^{52}$ erg in spherical symmetry as derived by modelling of their broad-lined spectra (Iwamoto et al. 1998; Nakamura et al. 2001; Mazzali et al. 2003; Deng et al. 2005; Mazzali et al. 2006a;b), larger than those of Type Ic narrow- and broad-lined SNe with no accompanying GRB (e.g., Mazzali et al. 2000; Mazzali et al. 2002; Sauer et al. 2006; Corsi et al. 2011; Valenti et al. 2008; Soderberg et al. 2010; Pignata et al. 2011; Taubenberger et al. 2006).  The inferred masses of their progenitors are in all cases larger than 20\,M$_\odot$.

At redshifts larger than 0.25,  and up to $z \approx 1$, a SN has been detected for many long GRBs in the form of a ``bump'' in the light curve of the optical counterpart, and a SN light curve has been derived by subtraction of the host galaxy and afterglow component (Galama et al. 2000; Bloom et al. 1999; Lazzati et al. 2001; Masetti et al. 2003; Bersier et al. 2006; Soderberg et al. 2006, 2007; Cano et al. 2011b; Cobb et al. 2010; Gorosabel et al. 2005).  However, both the distance and the uncertainty related to this subtraction limit the signal-to-noise ratio (S/N) of the spectra, so that SN spectral features can only be identified in a single spectrum taken close to maximum light (Della Valle et al. 2003, 2006ab, 2008; Greiner et al. 2003; Garnavich et al. 2003; Soderberg et al. 2005; Berger et al. 2011; Sparre et al. 2011).

The lack of a well-sampled multi-band light curve and of a spectral time sequence prevents the accurate determination of the model parameters and of the SN physical properties. A single-band luminosity of the SN at maximum is not a reliable proxy for the $^{56}$Ni mass, because it does not carry information on the evolving broad-band spectrum.  Similarly, a single spectrum acquired near the light-curve peak allows only a  rough estimate of the kinetic energy, because it samples only a very limited fraction of the ejecta, causing the uncertainty on the kinetic energy to reach 50\%. Only well-monitored bolometric light curves and time-resolved spectra can probe the physics of these SNe, but the construction of suitable datasets at $z > 0.2$ has proven to be very difficult.

The long GRB\,120422A has a low redshift ($z = 0.283$; Tanvir et al. 2012; Schulze et al. 2012), which allowed an extensive coordinated observational campaign to be undertaken. This led to the discovery, spectroscopic identification, and subsequent monitoring of the associated SN\,2012bz (Malesani et al. 2012c). Here we present our observations, covering more than one month, which were carried out with TNG and the VLT. Throughout the paper, we assume a standard cosmology with H$_{\rm 0}$ = 72 km s$^{-1}$ Mpc$^{-1}$, $\Omega_{\rm m}$ = 0.27, and $\Omega_{\rm \Lambda}$ = 0.73.

\begin{figure*}
   \centering
   \includegraphics[width=18.5cm,height=2.5cm]{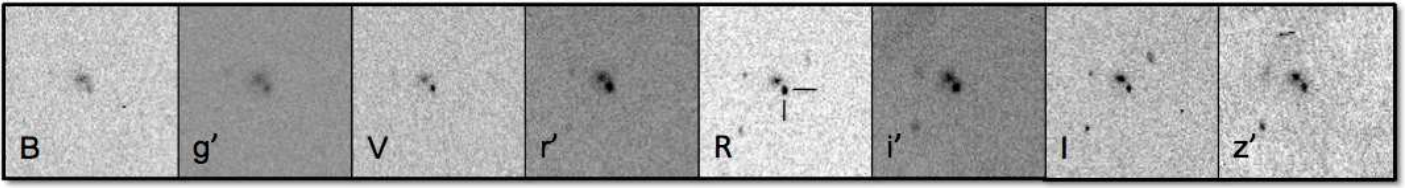}
      \caption{Images of SN\,2012bz  acquired at about 13.6 (VLT-$BVRI$) and 14.6 (TNG-$g'r'i'z'$) days after the burst. The position of SN\,2012bz, visible in all filters, is indicated  in the $R$-band stamp.}
         \label{FigSN}
\end{figure*}

\section{GRB\,120422A\,/\,SN\,2012bz}

On 2012 April 22 ($T_{0}$ = 07:12:03 UT) the Burst Alert Telescope (BAT) onboard the {\it Swift} spacecraft discovered GRB\,120422A, an event with a duration $T_{\rm 90} \approx 5$ s (Troja et al. 2012; Barthelmy et al. 2012). The X-ray Telescope (XRT) detected an uncatalogued X-ray fading source (Troja et al. 2012; Beardmore et al. 2012) also seen by the UV-Optical Telescope (Kuin et al. 2012) and by ground-based telescopes in the optical (Cucchiara et al. 2012; Schulze et al. 2012a; Nardini et al. 2012) and radio bands, at J2000 coordinates $\alpha$ = 09$^{h}$07$^{m}$38.42$^{s}$ ($\pm 0.01$), $\delta$ = +14$^{\circ}$01$^{\prime}$07.1$^{\prime\prime}$ ($\pm 0.2^{\prime\prime}$) (Zauderer et al. 2012). 

A redshift of 0.283 was first measured by Tanvir et al. (2012) for the candidate host galaxy and later confirmed for the optical counterpart of GRB\,120422A (Schulze et al. 2012b). The afterglow was faint at early times ($R \geq 18.4$\,mag after $\sim 7$\,min, Guidorzi et al. 2012; $i' \approx 19.0$ mag after $\sim 50$ min, Cucchiara et al. 2012) and decayed with a shallow power law ($\alpha_{\it i'} \approx 0.7$) up to 1 day after the burst (Cucchiara et al. 2012; Nardini et al. 2012). A smooth rebrightening observed in the $i'$ band 4 days after the burst was interpreted as the signature of the emerging SN (Malesani et al. 2012a). Spectroscopic confirmations of the SN came within a few days (Wiersema et al. 2012; Malesani et al. 2012b; Sanchez-Ramirez et al. 2012). The host galaxy was studied by Levesque et al. (2012).

We extracted the time-integrated spectrum as observed by BAT (see Section 4.4). As shown by Zhang et al. (2012), this event displayed a main pulse from $T_{\rm 0}-5$\,s to $T_{\rm 0}+25$\,s and a second episode of emission from $T_{\rm 0}+48$\,s to $T_{\rm 0}+65$\,s. The total spectrum over these two episodes is well fitted by a simple power law with index $\Gamma_{\rm BAT} = 1.91^{+0.37}_{-0.33}$ (90$\%$ c.l.; $\chi^{2}$/d.o.f. = 5.6/6).  Considering the narrow energy range of BAT and the fact that over a wider energy range GRB spectra are typically fitted with an empirical function (Band et al. 1993), we also tried to fit this function to the time-integrated spectrum of GRB\,120422A. We fixed its slopes to $\alpha=-1.0$ and $\beta=-2.3$ (typically observed in GRB spectra; e.g., Kaneko et al. 2006). The best-fit value of the peak energy is $E_{\rm p,i} = 33^{+39}_{-33} \leq 72$\,keV (90$\%$ c.l.) and the corresponding $E_{\rm iso}$ in the rest-frame 1\,keV -- 1\,MeV energy range is (1.6--3.2) $\times 10^{50}$ erg.

\section{Observations and Data Reduction}

We observed the field of GRB\,120422A with TNG equipped with DOLORES (imaging in the $g'r'i'z'$ filters) and with the VLT equipped with FORS2 (imaging in $BVRI$ and spectroscopy) from 4.5 to 35.7 days after the burst (see Fig. \ref{FigSN}). Tables 1 and 2 summarize our observations. 
 
\subsection{Imaging}

Image reduction, including de-biasing and flat-fielding, was carried out following standard procedures. Images were calibrated using a set of Sloan Digital Sky Survey (SDSS) stars acquired with SDSS $g'r'i'z'$ filters (TNG observations) and with respect to standard fields in $BVRI$ filters (VLT observations). We performed point-spread function (PSF) photometry at the position of the optical afterglow in order to minimize the possible contribution of the host galaxy.

\begin{figure}
   \centering
   \includegraphics[width=7.0cm,height=9.0cm,angle=-90]{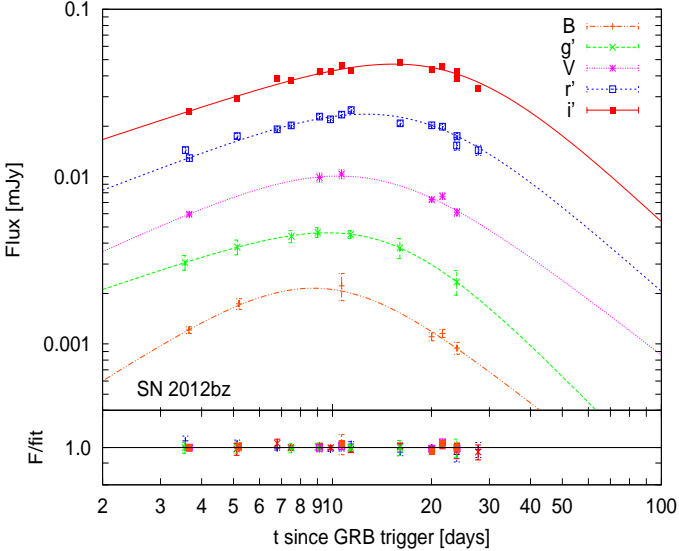}
      \caption{Multi-band light curves of SN\,2012bz in the rest frame. We converted the magnitudes into flux density (Fukugita et al. 1995), after taking into account the Galactic extinction. Fluxes in the {\bf $Bg'Vr'i'$} filters have been shifted for clarity by a factor 1, 1.0, 1.2, 2, and 4, respectively. Fits to the light curves with broken power laws are also shown.}
      \label{lcSN}
\end{figure}

\subsection{Spectroscopy}

VLT FORS2 spectroscopy was carried out using the 300V grism, covering the range 4250--9500\,\AA. We used in all cases a $1''$ slit, resulting in an effective resolution $R = 440$ at the central wavelength $\lambda = 5500\,\AA$. The spectra were extracted using standard procedures under the packages ESO-MIDAS\footnote{\texttt{http://www.eso.org/projects/esomidas/ .}} and IRAF\footnote{\texttt{http://iraf.noao.edu/ .}}. A He-Ar lamp and spectrophotometric stars were used to calibrate the spectra in wavelength and flux.  We accounted for slit losses by matching the flux-calibrated spectra to our multi-band photometry.

\begin{table*}
 \centering
 \caption[]{Observation log for GRB\,120422A\,/\,SN\,2012bz: $\Delta t$ refers to the beginning of the observations and corresponds to the delay since the burst event. $UBVRI$ magnitudes are in the Vega system, $g'r'i'z'$ magnitudes are in the AB system, and they have not been corrected for Galactic absorption. Errors are given at the 1$\sigma$ confidence level.}
 \label{logmag}
  \begin{tabular}{cccccc|cccccc}
\hline
\hline
$\Delta t$ & $t_{\rm exp}$ & Filt. & Mag & Flux & Instrument & $\Delta t$ & $t_{\rm exp}$ & Filt. & Mag & Flux & Instrument \\
\hline
(days) & (min) & & & ($10^{-3}$\,mJy) & & (days) & (min) & & & ($10^{-3}$\,mJy) & \\
\hline
11.714 & 2.0 & $U$ & $>$21.1 & $>$5.86 & VLT/FORS2                 & 14.619 & 10.0 & $r'$ & 21.25 $\pm$ 0.05 & 11.48 $\pm$ 0.53 & TNG/LRS \\
13.678 & 2.0 & $U$ & $>$21.0 & $>$5.86 & VLT/FORS2              	 & 20.610 & 10.0 & $r'$ & 21.45 $\pm$ 0.07 & 9.55 $\pm$ 0.61 & TNG/LRS \\
4.697 & 6.5 & $B$ & 23.46 $\pm$ 0.10 & 1.70 $\pm$ 0.15 & VLT/FORS2      	 & 30.608 & 10.0 & $r'$ & 21.78 $\pm$ 0.14 & 7.05 $\pm$ 0.90 & TNG/LRS \\
6,674 & 3.0 & $B$ & 23.07 $\pm$ 0.08 & 2.44 $\pm$ 0.18 & VLT/FORS2         & 35.612 & 15.0 & $r'$ & 21.85 $\pm$ 0.15 & 6.60 $\pm$ 0.91 & TNG/LRS \\
11.715 & 1.0 & $B$ & $>$ 22.3 & $>$4.97 & VLT/FORS2             	 &  4.712 & 3.0 & $R$ & 21.79 $\pm$ 0.02 & 5.97 $\pm$ 0.11 & VLT/FORS2 \\
13.680 & 1.0 & $B$ & 22.80 $\pm$ 0.20 & 3.13 $\pm$ 0.57 & VLT/FORS2     	 &  8.709 & 3.0 & $R$ & 21.36 $\pm$ 0.04 & 8.87 $\pm$ 0.32 & VLT/FORS2 \\ 
25.700 & 1.0 & $B$ & 23.56 $\pm$ 0.06 & 1.56 $\pm$ 0.09 & VLT/FORS2     	 &  11.719 & 1.0 & $R$ & 21.23 $\pm$ 0.08 & 10.00 $\pm$ 0.73 & VLT/FORS2\\
27.710 & 1.0 & $B$ & 23.51 $\pm$ 0.06 & 1.63 $\pm$ 0.09 & VLT/FORS2     	 &  13.684 & 1.0 & $R$ & 21.14 $\pm$ 0.08 & 10.87 $\pm$ 0.80 & VLT/FORS2\\
30.677 & 4.0 & $B$ & 23.73 $\pm$ 0.09 & 1.33 $\pm$ 0.11 & VLT/FORS2     	 &  25.707 & 1.0 & $R$ & 21.30 $\pm$ 0.03 & 9.37 $\pm$ 0.26 & VLT/FORS2 \\
4.568 & 5.0 & $g'$ & 22.82 $\pm$ 0.11 & 2.70 $\pm$ 0.27 & TNG/LRS   	  	 &  27.707 & 1.0 & $R$ & 21.32 $\pm$ 0.05 & 9.21 $\pm$ 0.42 & VLT/FORS2\\
6.593 & 5.0 & $g'$ & 22.69 $\pm$ 0.11 & 3.05 $\pm$ 0.30 & TNG/LRS   	      	 & 30.683 & 1.0 & $R$ & 21.46 $\pm$ 0.04 & 8.09 $\pm$ 0.29 & VLT/FORS2 \\
9.628 & 15.0 & $g'$ & 22.43 $\pm$ 0.09 & 3.87 $\pm$ 0.32 & TNG/LRS  	          & 6.558 & 5.0 & $i'$ & 21.80 $\pm$ 0.12 &  6.92 $\pm$ 0.76 & TNG/LRS \\      
11.577 & 15.0 & $g'$ & 22.37 $\pm$ 0.07 & 4.09 $\pm$ 0.26 & TNG/LRS 	 & 9.581 & 15.0 & $i'$ & 21.54 $\pm$ 0.05 & 8.79 $\pm$ 0.40 & TNG/LRS \\     
14.566 & 15.0 & $g'$ & 22.40 $\pm$ 0.06 & 3.98 $\pm$ 0.22 & TNG/LRS 	 & 12.670 & 10.0 & $i'$ & 21.40 $\pm$ 0.04 & 10.03 $\pm$ 0.37 & TNG/LRS \\    
20.562 & 10.0 & $g'$ & 22.60 $\pm$ 0.15 & 3.31 $\pm$ 0.45 & TNG/LRS 	 & 14.596 & 10.0 & $i'$ & 21.39 $\pm$ 0.05 & 10.09 $\pm$ 0.46 & TNG/LRS \\   
30.619 & 10.0 & $g'$ & 23.11 $\pm$ 0.18 & 2.07 $\pm$ 0.34 & TNG/LRS 	& 20.589 & 10.0 & $i'$ & 21.26 $\pm$ 0.06 & 11.38 $\pm$ 0.63 & TNG/LRS \\    
4.707 & 3.0 & $V$ & 21.89 $\pm$ 0.03 & 6.57 $\pm$ 0.18 & VLT/FORS2      	  & 30.596 & 10.0 & $i'$ & 21.51 $\pm$ 0.14 & 9.04 $\pm$ 1.16 & TNG/LRS \\   
11.718 & 1.0 & $V$ & 21.34 $\pm$ 0.08 & 10.74 $\pm$ 0.79 & VLT/FORS2     	 & 35.573 & 18.0 & $i'$ & 21.65 $\pm$ 0.15 & 7.94 $\pm$ 1.09 & TNG/LRS \\  
13.682 & 1.0 & $V$ & 21.29 $\pm$ 0.08 & 11.25 $\pm$ 0.82 & VLT/FORS2     	  &  4.707 & 3.0 & $I$ & 21.62 $\pm$ 0.06 & 5.47 $\pm$ 0.30 & VLT/FORS2 \\ 
25.706 & 1.0 & $V$ & 21.67 $\pm$ 0.03 & 7.93 $\pm$ 0.22 & VLT/FORS2     	  & 8.716 & 3.0 & $I$ & 21.21 $\pm$ 0.08 & 7.98 $\pm$ 0.59 & VLT/FORS2 \\ 
27.706 & 1.0 & $V$ & 21.62 $\pm$ 0.06 & 8.30 $\pm$ 0.46 & VLT/FORS2     	  & 11.721 & 1.0 & $I$ & 20.93 $\pm$ 0.06 & 10.33 $\pm$ 0.57 & VLT/FORS2 \\ 
30.682 & 1.0 & $V$ & 21.86 $\pm$ 0.07 & 6.65 $\pm$ 0.43 & VLT/FORS2     	 & 13.685 & 1.0 & $I$ & 20.84 $\pm$ 0.08 & 11.22 $\pm$ 0.83 & VLT/FORS2 \\
4.580 & 5.0 & $r'$ & 21.85 $\pm$ 0.10 & 6.60 $\pm$ 0.60 & TNG/LRS       	 & 25.708 & 1.0 & $I$ & 20.90 $\pm$ 0.05 & 10.62 $\pm$ 0.49 & VLT/FORS2\\
6.582 & 5.0 & $r'$ & 21.64 $\pm$ 0.08 & 8.02 $\pm$ 0.59 & TNG/LRS 	          & 27.709 & 1.0 & $I$ & 20.85 $\pm$ 0.07 & 11.12 $\pm$ 0.72 & VLT/FORS2 \\ 
9.606 & 15.0 & $r'$ & 21.48 $\pm$ 0.04 & 9.29 $\pm$ 0.34 & TNG/LRS	 & 30.685 & 1.0 & $I$ & 20.93 $\pm$ 0.05 & 10.33 $\pm$ 0.47 & VLT/FORS2 \\   
12.650 & 10.0 & $r'$ & 21.39 $\pm$ 0.04 & 10.09 $\pm$ 0.37 & TNG/LRS	 & 14.643 & 20.0 & $z'$ & 21.64 $\pm$ 0.08 & 8.02 $\pm$ 0.59 & TNG/LRS \\
\hline
 \end{tabular}
\end{table*}

\section{Results}

\subsection{Optical Light Curve}

The GRB optical counterpart brightened during our monitoring, reaching maximum generally at later times in redder filters (Fig. \ref{lcSN})\footnote{In this figure we decided to unify $R$ and $r'$ ($I$ and $i'$) data under a single label $r'$ ($i'$).}. A simple estimate of the peak time for each filter, fitting the rest-frame light curve with a generic broken power-law function, provides the following values: $t_{\rm peak}(B) = 9.4 \pm 3.8$, $t_{\rm peak}(g') = 14.5 \pm 0.7$, $t_{\rm peak}(V) = 12.9 \pm 4.3$, $t_{\rm peak}(r') = 17.6 \pm 2.4$, and $t_{\rm peak}(i') = 23.2 \pm 2.9$ days. This flux increase, the epoch of maximum, and its wavelength dependence suggest that the source is dominated by a SN component. The decaying phase after 20 days appears to be achromatic. Lacking data at epochs earlier than 4.5 days, we cannot accurately quantify the contribution of the GRB afterglow to the SN light.

\subsection{Optical Spectra}

The optical spectra of GRB\,120422A\,/\,SN\,2012bz have been cleaned of host-galaxy emission lines as well as telluric features and smoothed with a boxcar of 10\,\AA.  No correction was applied for intrinsic extinction, which appears to be negligible since the GRB counterpart was detected in all of the UV filters (Kuin et al. 2012). A correction for Galactic extinction was first applied ($E_{B-V}=0.037$\,mag; Schlegel, Finkbeiner, $\&$ Davis 1998). Then, we calculated the host-galaxy continuum contribution at the position of the SN, that is located in an outer region of the host galaxy, at a projected offset of $\sim 8$ kpc from its center. The galactic background in that position was found to be approximately an order of magnitude less than the integrated emission of the whole galaxy.

\begin{figure}
   \centering
   \includegraphics[width=9.0cm,height=7.5cm]{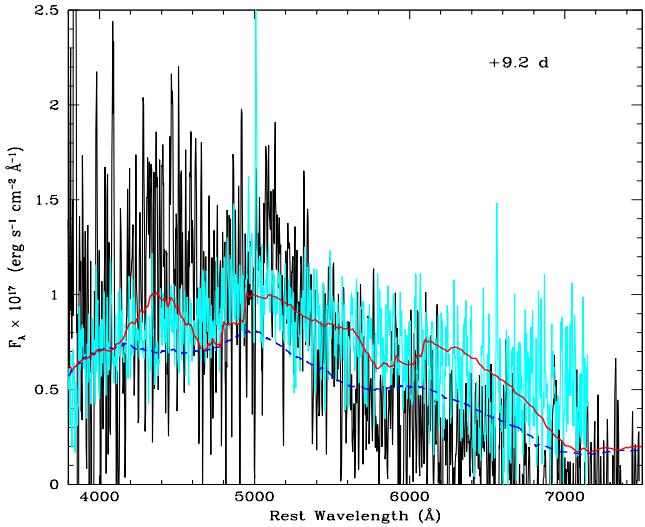}
      \caption{FORS2 spectrum of SN\,2012bz in the rest frame (black), smoothed with a boxcar of 10\,$\AA$, cleaned of spurious features and host-galaxy emission lines, and corrected for Galactic extinction, compared to the reddening-corrected and rescaled spectra of SN\,1998bw (Patat et al. 2001; blue dashed line) and SN\,2003dh (Mazzali et al. 2003; Deng et al. 2004; cyan) at comparable rest-frame phases after explosion, and also with models for SN\,2003dh (Mazzali et al. 2003; red solid line).}
     \label{SN_spec1}
\end{figure}

\begin{figure*}
   \centering
   \includegraphics[width=16.0cm,height=14.0cm]{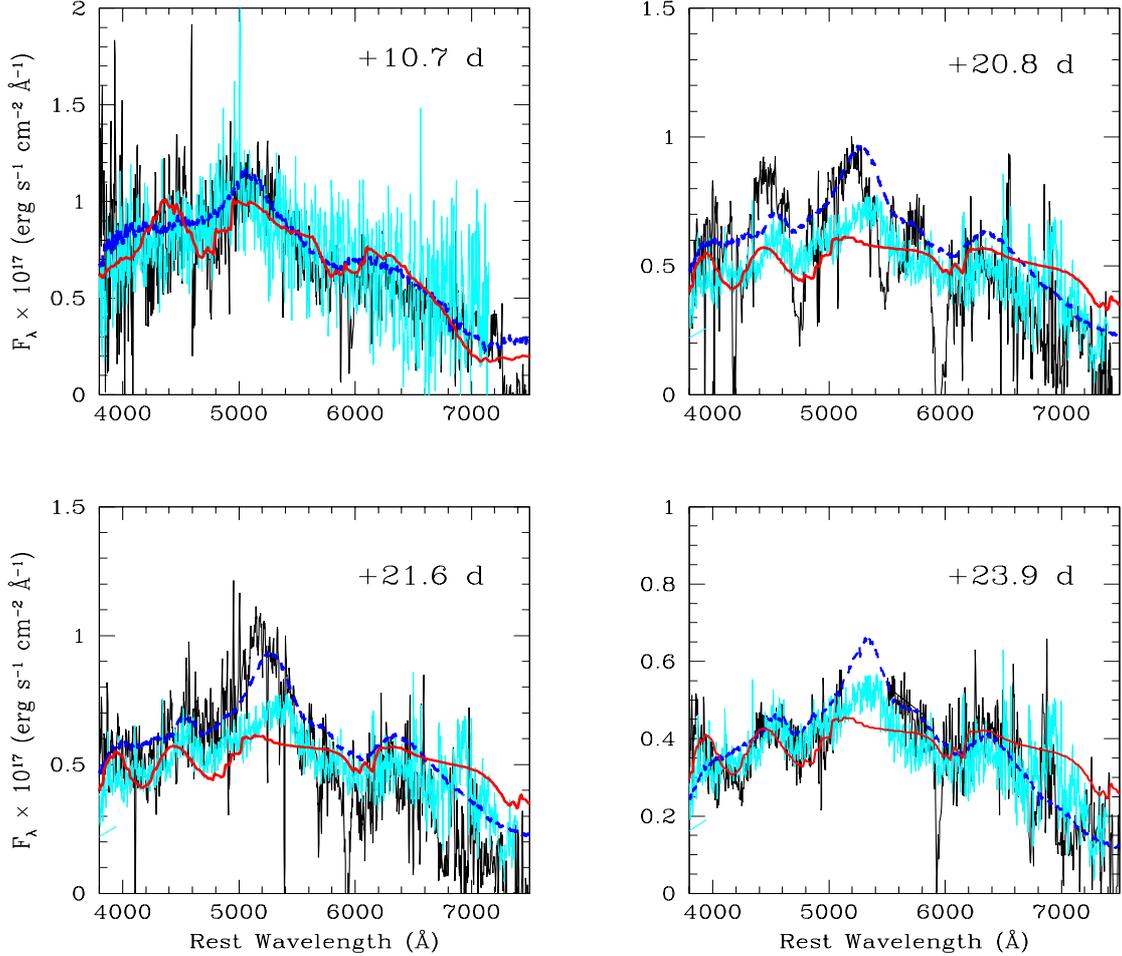}
      \caption{Same as Fig. \ref{SN_spec1} for epochs +10.7, +20.8, +21.6, and +23.9 rest-frame days after explosion.}
     \label{SN_spec}
\end{figure*}

In Fig. \ref{SN_spec1} we show our first spectrum, acquired 9.2 days (rest frame) after the burst event, when SN\,2012bz was still emerging and had not yet reached its maximum of emission, while in Fig. \ref{SN_spec} the next four spectra of our campaign are displayed. All the spectra are compared to those of SN\,1998bw (blue) and SN\,2003dh (cyan) at comparable rest-frame phases after explosion. We also overplot the spectral models (red) of Mazzali et al. (2003) for SN\,2003dh. The FORS2 spectra of SN\,2012bz are rather noisy at $\lambda_{\rm rest-frame} < 3800\,\AA$ and $\lambda_{\rm rest-frame} > 7400\,\AA$; thus, we present them only between these two wavelengths. Their quality is not sufficient for the accurate measurement of absorption features, so we cannot estimate photospheric velocities.  However, they present convincing similarity with spectra of SN\,1998bw and with the spectral models of  SN\,2003dh, within the noise.

\begin{table}
\centering
\caption{Summary of VLT/FORS2 spectroscopic observations of GRB\,120422A\,/\,SN\,2012bz: $\Delta t$ corresponds to the starting acquisition time of the spectra in the observed frame, and the average S/N is calculated per spectral bin.}
\begin{tabular}{ccccc}
\hline
$\Delta t$ & Seeing & $<$S/N$>$ & $t_{\rm exp}$ & Grism \\
 \hline
 (days) & (arcsec) & & (s) & \\
   \hline
11.776 & 0.9 & $\sim$3 & 1 $\times$ 1800 & 300V \\
13.690 & 0.6 & $\sim$8 &2 $\times$ 1800 & 300V \\
26.683 & 1.2 & $\sim$5 & 2 $\times$ 1800 & 300V \\
27.727 & 1.1 & $\sim$11 & 2 $\times$ 1800 & 300V \\
30.705 &  1.0 & $\sim$11 &2 $\times$ 1800 & 300V \\
\hline
\end{tabular}
\label{tab_log_spe}
\end{table}

\subsection{Bolometric Light Curve}

As customarily done for SNe in order to estimate their ultraviolet-optical-infrared (UVOIR) bolometric light curves, we constructed quasi-simultaneous (within hours) optical broadband spectral distributions at 10 epochs using the available photometry corrected for Galactic extinction with $E_{B-V}=0.037$ mag, $R = 3.1$, and the extinction curve of Cardelli et al. (1989). As was done for the spectra (Fig. \ref{SN_spec}), we have corrected the photometric points for the host-galaxy flux estimated as explained in Section 4.2. The galaxy template flux was first averaged over 200\,\AA-wide bands centered on the filter central wavelengths and then subtracted. At any given epoch, the dereddened and galaxy-subtracted photometric points were interpolated and extrapolated through a spline function to cover the observed wavelength range 4250--9500\,\AA. 

The bolometric light-curve points were then obtained by integrating the spectral flux at each epoch on this interval. Their statistical uncertainties are the quadrature sums of the statistical errors in the observed photometry and in the standard deviations of the averaged template galaxy fluxes. However, a more significant source of error (a factor 3 to 17 larger than the statistical uncertainty) affecting the bolometric light curve is the inaccuracy of the galaxy subtraction (see discussion in Section 4.2) and of the variable afterglow component. The latter is very difficult to estimate because of the scarce photometric information at early epochs. 

\begin{figure}
   \centering
   \includegraphics[width=9.0cm,height=7.5cm]{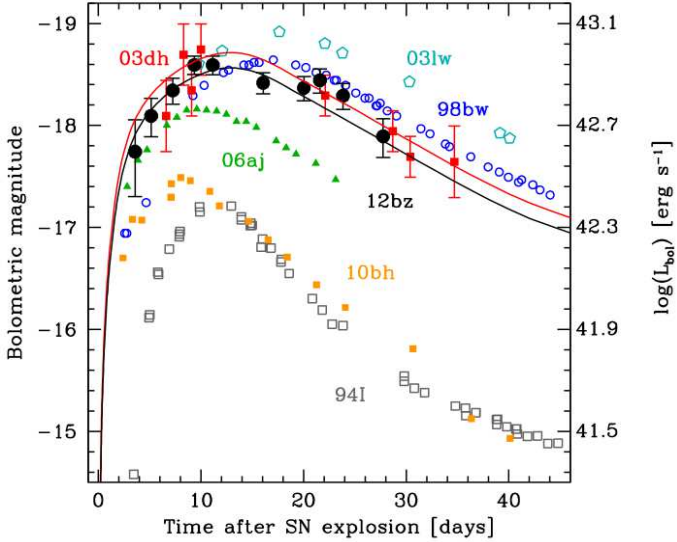}
      \caption{Bolometric light curves of GRB-SNe, XRF-SNe, and SN\,1994I in the corresponding rest frames, dereddened for Galactic extinction. When relevant, a correction for intrinsic extinction was also applied. The curve of SN\,2012bz (black points) is from the present work. The others were given by Pian et al. (2006) and Bufano et al. (2012). For clarity, errors in individual datapoints have been reported only for SN\,2012bz and SN\,2003dh. The model for SN\,2012bz (black line) was obtained by scaling down the model of SN\,2003dh (red line; Deng et al. 2005) by 15\%.}
     \label{lcSN_bol}
\end{figure}

Based on the photometric observations taken around 16.5\,hr after the GRB trigger time (Schulze et al. 2012a; Nardini et al. 2012; Rumyantsev et al. 2012; Malesani et al. 2012), when the SN contribution was negligible, we constructed an afterglow spectral energy distribution (SED), and assumed that the flux should decay thereafter to a level that, 4 days after the GRB, does not exceed that of the SN. This minimum decay rate would correspond to an index $\alpha \approx 0.85$ ($f(t) \propto t^{-\alpha}$). Assuming this decay rate and a maximum brightness for the putative host galaxy, we obtain an estimate of the minimum allowed flux for SN\,2012bz. If, on the other hand, we assume no significant contributions from the galaxy and the afterglow, we obtain an estimate of the maximum SN flux. The bolometric flux points reported in Fig. \ref{lcSN_bol} are the averages of these extreme estimates. The uncertainties represent the extent of the two extreme estimates, summed in quadrature with the statistical errors.

As in the case of the spectra, we find a similarity between the light curve of SN\,2012bz and that of the GRB-associated SN\,2003dh. We then applied to the bolometric light curve the model constructed by Deng et al. (2005) for SN\,2003dh, scaled down by 15\%. This represents a good fit of the curve of SN\,2012bz. The correspondingly scaled amount of synthesized $^{56}$Ni is thus predicted to be about 0.35~M$_{\odot}$, with other model parameters (e.g., ejecta masses) being unchanged (the scaling factor is sufficiently small that we can assume that the model would still return a self-consistent result).

The bolometric light curve of SN\,2012bz is compared in Fig. \ref{lcSN_bol} with those of other GRB-SNe and XRF-SNe, and with that of the normal Type Ic SN\,1994I. The rest-frame wavelength interval over which the bolometric light curve of SN\,2012bz was computed is 3300--7400\,\AA. For GRB-SNe at lower redshifts, this rest-frame interval extends to about 9000\,\AA, and a near-IR correction is included (this generally does not exceed 15$\%$ in SNe of this type). We preferred to avoid making arbitrary corrections for the missing flux in the red and near-IR part of the spectrum of SN\,2012bz, since this would introduce further inaccuracies. Thus, the present light curve should be conservatively considered as a lower limit to the actual one.

Although the rest-frame bolometric light curve reaches its maximum after $13 \pm 1$ days, we note that it seems to exhibit a rebrightening with respect to the best-fit model after 20 rest-frame days. Since we see no evidence for this in the monochromatic light curves (Fig. \ref{lcSN}), we conclude that the discrepancy has no physical origin; it may rather be due to a calibration error. The observed flux excess with respect to the model is about 25\%, only a factor of 2 larger than the uncertainty associated with those two points.

\begin{figure}[h]
   \centering
   \includegraphics[width=9.5cm,height=7.5cm]{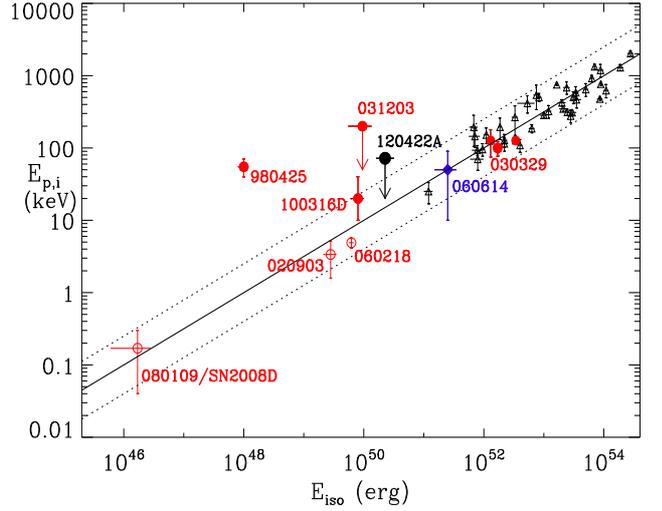}
      \caption{Intrinsic peak energy of long GRBs versus energy output in the rest-frame 1\,keV -- 1\,MeV band. GRB\,120422A is plotted as a black circle. The red filled (open) circles are other GRBs (XRFs) with associated SNe. The two energetic events close to GRB\,030329 are GRB\,051221 and GRB\,021211, with names omitted for clarity. As blue diamonds we show GRB\,060614, one of the two nearby, long GRBs without associated SNe (we omitted GRB\,060505 since it is not a bona-fide long GRB). We show the 2$\sigma$ confidence levels of the correlation with dotted lines.}
     \label{epeiso}
\end{figure}

\subsection{$E_{\rm p,i}-E_{\rm iso}$ Consistency}

With only the exception of GRB\,980425, the original discovery case of the connection between GRBs and SNe (Galama et al. 1998), long GRBs appear to obey an empirical power-law relation ($E_{\rm peak} \propto E_{\rm iso}^{0.5}$) between their isotropically estimated  prompt emission in the 1 keV -- 1 MeV range, $E_{\rm iso}$, and the peak energies of their spectra, $E_{\rm peak}$ (Amati et al. 2002, 2007, 2012). In this relationship, which is reported in updated form in Fig. \ref{epeiso}, XRFs, the softest variety of long GRBs, occupy the lowest isotropic-energy tail of the distribution.

To verify the consistency of this burst with the $E_{\rm p,i}-E_{\rm iso}$ correlation (Amati et al. 2002), we extracted the entire time-integrated spectrum observed by BAT (Sec. 2). Fig. \ref{epeiso} shows that the value of $E_{\rm p,i}$ is very similar to those of other GRB-SNe, including the major outlier of the correlation (GRB\,980425), for which possible explanations have been proposed (Ramirez-Ruiz et al. 2005; Ghisellini et al. 2006). The data available for the prompt emission of GRB\,120422A might not be good enough to firmly establish an inconsistency with the $E_{\rm p,i}-E_{\rm iso}$ correlation. In fact, the upper limit for its value of $E_{\rm p,i}$ (similarly to GRB\,031203) makes this event consistent with the correlation. We note that GRB\,120422A has an X-ray total energy in the rest-frame 0.3--30\,keV energy range of $E_{\rm X,iso} = (5.49 \pm 0.33) \times 10^{49}$\,erg, making this event consistent (at 1$\sigma$ c.l.) with the correlation found by Margutti et al. (2012) and Bernardini et al. (2012).


\section{Conclusion}

The rest-frame peak energy of the integrated spectrum of GRB\,120422A, lower than or equal to 70\,keV, makes it a possible representative of the X-ray-rich class, intermediate between GRBs and XRFs. Its associated supernova, SN\,2012bz, may therefore be a link between GRB-SNe and XRF-SNe, although its properties seem more similar to those of GRB-SNe than to those of XRF-SNe. The upper limit on the GRB peak energy and the SN maximum luminosity of GRB\,120422A\,/\,SN\,2012bz are consistent with the empirical relationship of Li (2006). The wide range of total energies emitted by the gamma-ray and X-ray events (4 orders of magnitude), as opposed to the much narrower span of maximum luminosities of the respective SNe (a factor of 3), suggests that the properties of the GRBs/XRFs are a strong function of the viewing angle (Guetta \& Della Valle 2007; Podsiadlowski et al. 2004).

GRB\,120422A\,/\,SN\,2012bz exhibits the GRB-SN connection at the highest redshift so far accurately monitored both photometrically and spectroscopically. The maximum luminosity of SN\,2012bz and the shape of its bolometric light curve are very similar to those of the three previously known SNe associated with GRBs at $z < 0.2$ (SN\,1998bw, SN\,2003dh, and SN\,2003lw). In particular, a comparison of both spectra and light curves with those of SN\,2003dh (Deng et al. 2005; Mazzali et al. 2006a) show a close match, and therefore comparable explosion conditions and parameters. This leads us to suggest for SN\,2012bz a mass of synthesized $^{56}$Ni of $\sim 0.35$\,M$_{\odot}$ (about 15\% smaller than that of SN 2003dh), a similar kinetic energy ($E_{\rm K} \approx 3.5 \times 10^{52}$ erg), similar ejecta mass ($M_{\rm ej} \approx 7\,{\rm M}_{\odot}$), and, consequently, a comparable mass of the progenitor star, 25--40\,M$_{\odot}$. More details on the geometry of the explosion can be obtained with spectroscopy in the nebular phase, 6--12 months after the explosion.

\begin{acknowledgements}

We thank the anonymous referee for valuable comments and suggestions that improved the paper. We thank the TNG staff, in particular W. Boschin, M. Cecconi, L. di Fabrizio, F. Ghinassi, A. Harutyunyan, and M. Pedani, for their valuable support with TNG observations, and the Paranal Science Operations Team, in particular H. Boffin, S. Brillant, D. Gadotti, D. Jones, M. Rodrigues, L. Schmidtobreick, and J. Smoker. We are grateful to D. Malesani, J. Fynbo, N. Tanvir, and K. Wiersema for their support with VLT observation planning.  We acknowledge support from PRIN INAF 2009 and 2011, PRIN-MIUR grant 2009ERC3HT and from grants ASI INAF I/088/06/0 and I/011/07/0. E. Pian is grateful for hospitality at the ESO Headquarters in Santiago, where part of this work was developed. F.B. acknowledges support from FONDECYT through Postdoctoral grant 3120227. J.D. is supported by the 973 Program of China (Grant No. 2009CB824800). R.L.C.S. is supported by a Royal Society Fellowship. A.V.F. is grateful for the support of US NSF grants AST-0908886 and AST-1211916, the TABASGO Foundation, and the Christopher R. Redlich Fund.

\end{acknowledgements}

\end{document}